\documentclass[aps,prb,preprint]{revtex4}
\usepackage{amsmath,amssymb,epsfig}

\begin{document}


\title{Correlation and confinement induced itinerant ferromagnetism in chain 
structures}

\author{R\'eka~Trencs\'enyi$^{a}$, Endre~Kov\'acs$^{b}$, and  
Zsolt~Gul\'acsi$^{a}$}

\address{$^{(a)}$ Department of Theoretical
Physics, University of Debrecen, H-4010 Debrecen, Hungary\\
$^{(b)}$ Department of Physics, University of Miskolc, H-3515 
Miskolc-Egyetemvaros, Hungary}

\date{June 1, 2009}

\begin{abstract}
Using a positive semidefinite operator technique one deduces exact ground 
states for a zig-zag
hexagon chain described by a non-integrable Hubbard model with on-site 
repulsion. 
Flat bands are not present in the bare band structure, and the operators 
$\hat B^{\dagger}_{\mu,\sigma}$ introducing the electrons into the ground 
state, are all extended operators and confined in the quasi 1D
chain structure of the system. Consequently, increasing the number of carriers,
the $\hat B^{\dagger}_{\mu,\sigma}$ operators become connected 
i.e. touch each other on several lattice
sites. Hence the spin projection of the carriers becomes correlated in 
order to minimize the ground state
energy by reducing as much as possible the double occupancy leading to a
ferromagnetic ground state. This
result demonstrates in exact terms in a many-body frame that the conjecture 
made at two-particle level
by G. Brocks et al. [Phys.Rev.Lett. {\bf 93},146405, (2004)] that the Coulomb 
interaction is expected to 
stabilize correlated magnetic ground states in acenes is clearly viable, and 
opens new directions 
in the search for routes in obtaining organic ferromagnetism. Due to the 
itinerant
nature of the obtained ferromagnetic ground state, the systems under 
discussion may have also direct application
possibilities in spintronics.  
\end{abstract}

 \maketitle


\section{Introduction}

\subsection{Conducting chains}

Conducting chains have been extensively studied in the past period, due to
several 
reasons. First, periodic chains holding different type of cells are genuine 
objects for applications in device design techniques at the level of molecular
nanotechnology \cite{BB1}. As such, these systems represent a starting point
in creating functional devices, materials and components on a 1-100 nanometer
length scale which can lead to new routes in realizing functions of practical
interest like field effect transistors, electroluminiscent diodes, 
nanocatalysis, hydrogen storage, etc. \cite{BB2}. Second, several type of 
periodic
conducting chains are in fact organic materials. Their nature, besides the fact
that potentially allows developments that can lead to condensates on a plastic 
(as plastic ferromagnetism for example \cite{BB2a}), has exceptional 
qualities for 
new developments in electronics, for example in the direction of organic 
optoelectronic and field effect 
transistor components \cite{BB3}, high performance transistors and circuits
made based on soluble materials \cite{BB4}, biodegradable devices used for 
controlled-release drug delivery inside of the human body \cite{BB5}, or
electronic components on plastic providing flexible electronics \cite{BB6}.
Third, conducting chains exhibit quite interesting properties and phases 
important in advanced technological applications (as spintronics), for 
example insulating, conducting, half metallic, paramagnetic or ferromagnetic 
behavior \cite{BB7}. Furthermore, these systems have the virtue 
of having fundamentally different ground states which can be tuned by
external parameters like external fields or site 
selective gate potentials \cite{BB8}, opening new routes for the design of 
valves, switches, or control devices. Finally, the majority of conducting 
chains being 
described by non-integrable models, represent genuine challenges to theory
and necessitate the development of rigorous techniques to describe
the physical properties of these systems.

The chains under consideration are built up in fact from periodic arrangements
of rings. 
Their theoretical study goes back to
middle nineties and starts with triangular chains \cite{BB9a,BB9b,BB9c}. 
This case
attracted attention especially by the study of the emergence possibilities of
ferromagnetism in such systems
\cite{BB10a,BB10b}, analysis of the stability of this phase \cite{BB10c}, 
low temperature 
thermodynamics \cite{BB10d}, numerical studies in two band cases 
\cite{BB10e}, or even study of exact ground states in non-integrable situations
\cite{BB8}. 

The attempts to characterize the physical properties of quadrilateral chains 
have been started later, being intensified by the study of Aharonov-Bohm cage
properties \cite{BB11}, and have led to the description of quite interesting
ground state characteristics in the diamond chain case which are tunable by 
external fields providing for example correlated half metal behavior 
applicable for spin-valve design \cite{BB7}. 

The study of chains of pentagons intensified after 2000, when
A. Heeger, A. MacDiarmid and H. Shirakawa obtained the Nobel prize in 
chemistry for the discovery and development of conducting polymers. The most 
common representatives of this class of materials are based on polythiophene,
polypyrrole and polytriazole, all containing pentagon rings. The study of the
physical properties of these systems has been concentrated on the search for
flat band ferromagnetism \cite{BB2a,BB12a,BB12b}, and also the analysis of
their ground states in exact terms has been started \cite{BB12c}.

Following the interest in increase in the number of lattice sites
in the ring which forms the basis of the periodic chain,
one arrives at chains of hexagons  
which are the subject of this paper. The hexagon chains are of real
interest since are relatively abundant in nature. The main representative
of the zig-zag hexagon chains at the level of organic materials, are the
polyacenes. Acenes, or polyacenes are a class of polycyclic aromatic 
hydrocarbons made up of linearly fused hexagon rings. These systems attracted
attention because of optoelectronic and electrical engineering 
application possibilities which have led even to organic field effect 
transistors in which pentacene is incorporated \cite{BB3}. Furthermore, such 
chains present
potential possibilities for the design of soluble
acene-based transistors and circuits \cite{BB4}, tetracene and pentacene 
have been used
for the design of light emitting devices \cite{BB4a}, etc. Due to these
properties, the polyacene structures have been studied extensively; 
the theoretical studies have been 
mostly confined to mean-field type of descriptions \cite{BB4b}. 
One notes that the interest in hexagonal chains and structures is not 
restricted exclusively to carbon based
materials, since also other compounds present similar cell structure, for 
example boron-nitride ionic honeycomb systems \cite{BB4c}, etc.  

\subsection{The aim and technique used}

Recently, based on first 
principle calculations,  the Coulomb interaction has been analyzed in acenes
between two particles. It was shown \cite{BB13}  that the average Coulomb 
repulsion in
these systems can reach values around $U_{eff} \sim 4-5 eV$, the ratio of
$U_{eff}$ to the bandwidth $W$ being of order $U_{eff}/W \sim 2.3 - 7.6$.
These results show that correlation effects are very prominent in this type
of chains, leading the authors to the conjecture that by increasing the 
concentration of carriers in these systems, the Coulomb repulsion is expected
to stabilize correlated magnetic ground states.

The present paper has the aim to analyze this conjecture at a 
genuine many-body 
level. In order to treat the correlations properly and create 
premises for valuable conclusions, the study is made here at exact level.

Exact results for the chains under consideration, taking into account
their quasi 1D character and non-integrable nature, are extremely rare. One 
only knows exact ground states for triangle \cite{BB8}, and rhombus \cite{BB7}
cases. The method used allows the exact ground state to be deduced under
general circumstances independent on dimensionality or integrability and it 
is based on a technique which uses positive semidefinite operator 
characteristics.
A positive semidefinite operator $\hat P$, is the operator which has the 
property $\langle \phi |\hat P | \phi \rangle \geq 0$ for all vectors 
$|\phi \rangle$ of the Hilbert space ${\cal{H}}$. It results that all 
eigenvalues $p_i$ of $\hat P$, $\hat P |\phi_i\rangle = p_i |\phi_i\rangle$,
are non-negative, e.g. $p_i \geq 0$. Consequently, if one succeeds 
in finding an exact decomposition of the
Hamiltonian $\hat H$ in terms of positive semidefinite terms, namely
$\hat H = \sum^m_n \hat P_n + C$, where $\hat P_n$ are positive semidefinite 
operators (whose number is $m$), and $C$ is a constant depending on 
Hamiltonian parameters, then the operator $\hat H'= \hat H - C$ has 
a positive
semidefinite form. Hence, $\hat H'$ has a spectrum bounded below by a well 
defined and known number, namely zero. 
As a result, the exact ground state
of $\hat H'$ (consequently, also of $\hat H$), is given by the vector $|\Psi_g
\rangle$ holding the property $\hat H' |\Psi_g\rangle =0$, i.e. 
$\hat P_n |\Psi_g\rangle =0$ must be satisfied for all $n=1,2,...,m$. 

The technique which we used is based on 
the above presented properties. First one has finds an exact decomposition of
the Hamiltonian of the system in a positive semidefinite form, and obtain 
the explicit form of the positive semidefinite operators
$\hat P_n$ for all $n$.
In the second step one constructs the $|\Psi_g \rangle$ ground state
Hilbert space vector such to satisfy $\hat P_n |\Psi_g\rangle = 0$ for
all $n$. This procedure depends on the explicit form of the $\hat P_n$
operators, but one exemplifies below a particular case which will be used
for the study of the zig-zag hexagon chain described below. 

Let us assume that one has for $n=1,2,...m_1 < m$ the structure
$\hat P_n=\hat A^{\dagger}_n \hat A_n$ for the positive semidefinite
operators $\hat P_n$, $n \leq m_1$, where 
$\hat A_n$ are built up from a linear combination of 
annihilation fermionic operators. In these conditions the construction of the
ground state starts by the construction of the wave vector
$|\Psi\rangle = \prod_{\mu}
\hat B^{\dagger}_{\mu} |0\rangle$ where $|0\rangle$ is the bare vacuum. 
The $\hat B^{\dagger}_{\mu}$ operators (constructed from fermionic creation
operators), are objects which have to be deduced. Their calculation is made
based on the anti-commutation relation $\{\hat A_n, \hat B^{\dagger}_{\mu} \} 
=0$, which must be satisfied for all values of all indices $n$ and $\mu$. 
Indeed, if this anti-commutation relation holds, in
$\hat P_n |\Psi\rangle = \hat A^{\dagger}_n \hat A_n (\prod_{\mu}
\hat B^{\dagger}_{\mu}) |0\rangle$, the $\hat A_n$ operator can be pushed in
front of the vacuum state obtaining $\hat P_n |\Psi\rangle = e^{i\phi}
\hat A^{\dagger}_n (\prod_{\mu}\hat B^{\dagger}_{\mu}) \hat A_n |0\rangle$,
which, given by the annihilation nature of $\hat A_n$ provides
$\hat A_n |0\rangle =0$. Consequently indeed  $\hat P_n |\Psi\rangle = 0$
holds. The phase factor $e^{i\phi}$ provides only a $+1$, or $-1$ 
multiplicative factor depending on the even, or odd number of operators in
$(\prod_{\mu} \hat B^{\dagger}_{\mu})$.
 
After deducing all possible $\hat B^{\dagger}_{\mu}$ operators, the 
ground state $|\Psi_g\rangle$ is obtained by restricting the index
$\mu \in I$, such to obtain $|\Psi_g\rangle = [ \prod_{\mu \in I}
\hat B^{\dagger}_{\mu}] |0\rangle$ based on the condition 
$\hat P_n |\Psi_g\rangle=0$ also for $n > m_1$. The corresponding ground state
energy becomes $E_g=C$. The carrier concentration at which $|\Psi_g\rangle$ is
defined, is provided by the number of electrons introduced into the system 
by the
$( \prod_{\mu \in I}\hat B^{\dagger}_{\mu})$ operator product acting on the
vacuum state $|0\rangle$.

One further notes that the transformation of the starting Hamiltonian $\hat H$
in a positive semidefinite form  $\sum^m_n \hat P_n + C$ is not unique, can be
performed in several different ways, each transformation places the final 
result
in different regions of the parameter space. The parameter space domain in 
which
the deduced ground state is present, is fixed by the matching conditions. These
last are relations between the $\hat H$ parameters and the $\hat P_n$ 
parameters
which allow the transcription of the starting $\hat H$ in the used positive
semidefinite form. 

The detailed presentation of the technique can be found in several recent
publications \cite{BB7,BB8,BB14,BB15}. The implementation of this method 
at finite value of the interaction has been
started at the end of nineties \cite{BB16,BB17,BB18,BB19,BB20,BB20a} and has
proven to be a successful and powerful technique
leading to exact
results even in situations unexpected in the context of exact solutions as:
three dimensions \cite{BB14,BB15}, disordered and interacting systems in 2D
\cite{BB21}, emergence of condensates \cite{BB21a,BB21b}, 
stripes and checkerboards in 2D \cite{BB22}, or insulator to 
metal transition driven by the Hubbard repulsion in 2D  in vicinity of half 
filling \cite{BB23}. We further note that similar techniques are used for
spin models as well \cite{BB23x}.

\subsection{Overview of the obtained results}

One analyzes the problem by concentrating on a fixed and given hexagon chain 
of zig-zag type (see Fig.1).
The zig-zag nature must be accentuated since hexagon cells can be connected 
in a chain also in
another way, namely in armchair configuration. The starting Hamiltonian, 
besides several hopping
matrix elements and on-site one-particle potentials contains also the Coulomb 
repulsion as interaction,
but for simplicity only at on-site level, providing the Hubbard interaction
term with $U > 0$ strength.
Since we are interested in finding valuable information about electron 
correlation effects in a fixed confined
system, one keeps the chain structure unchanged during the study, e.g. 
phononic contributions are 
neglected. Since Peierls transitions caused by electron-phonon interactions 
in chain structures with
even number of sites per cell could influence the emerging phases only 
around half-filling \cite{BB12a,BB12b}, 
one does not expect that the neglected phononic contributions will 
provide genuine
changes in our results
relating far from half-filling regions. 

The Hamiltonian of the starting chain is transformed first in a positive
semidefinite form. This transformation is important not only in the context of
acene structures. This is because several other systems of large interest 
today are
built up from hexagons. For example graphene, being a 2D hexagon structure 
constructed
from the same cell, can be described at the level of the transformation into a
positive semidefinite form by the same block operators as used here for the 
zig-zag hexagon chain 
in Section IV. The block operators applied for the transformation were such 
chosen to
provide a not severely restricted phase diagram region where the conclusions 
are valid.

After this
transformation one analyzes the bare band structure and one shows 
that
flat bands are not possible to emerge. Hence, flat band ferromagnetism 
\cite{BB10a,BB10b}
(often studied in the
context of conducting chains, see for example \cite{BB2a,BB12a,BB12b}) 
is excluded {\it a priori} from the spectrum of possible magnetic phases.

After the positive 
semidefinite form of the Hamiltonian has been obtained, 
one turns to
the construction of the ground states $|\Psi_g\rangle$. This step starts 
with
the deduction of the operators denoted hereafter by $\hat B^{\dagger}_{\mu}$, 
building
up $|\Psi_g\rangle$. One notes that the label index $\mu$  has also
a spin component.
Interestingly, it turns out that all these operators are extended,
i.e. have components placed in each cell and cannot be introduced in a 
restricted
domain of the chain with extension smaller than the system size in the 
direction
of the primitive vector. The extended nature of all $\hat B^{\dagger}_{\mu}$
operators is a special characteristic of the hexagon chains, 
which has not been
observed previously in the case of triangle, quadrilateral and pentagon chains
\cite{BB7,BB8,BB12c}. This aspect is important for two reasons. First,
the methods for the treatment and deduction of the extended operators in the
context of the positive semidefinite operator technique are 
practically completely open and unknown. 
On this line one notes that even if it has been shown that
the treatment in another context of the
extended operators is feasible \cite{xxx}, one
has only one case described 
quite recently in the literature, namely in the study of
the insulator to metal transition driven by the Hubbard repulsion in 
2D in the vicinity of half filling \cite{BB23}. Since a strategy developed
for 2D, directly cannot be applied for quasi 1D structures, at this point a
special technique has been constructed and described for the deduction of
the $\hat B^{\dagger}_{\mu}$ operators, which can be used for other chain 
structures as well. Second, the extended nature of $\hat B^{\dagger}_{\mu}$
terms is present in a confined space region delimited by the chain itself.
Consequently, by increasing the carrier concentration (i.e. increasing the
number of independent $\hat B^{\dagger}_{\mu}$ operators in $|\Psi_g\rangle$),
the $\hat B^{\dagger}_{\mu}$ operators present in the ground state wave vector
will satisfy connectivity conditions (i.e. will act on common lattice sites).
Given by this, the spin indices of the electrons must be correlated in order to
avoid as much as possible the double occupancy and to reduce in this manner
the ground state energy. This is the route for the emergence of ferromagnetism
in these systems. The obtained results show that the conjecture 
by Brocks et al.
\cite{BB13} made at two-particle level for acenes is indeed viable in a
rigorous many-body frame provided by a fixed zig-zag hexagon chain described
by a Hubbard type of model.

The remaining part of the paper is structured as follows: Section II. 
describes in detail the system of interest and its Hamiltonian, Section III.
analyzes the non-interacting band structure, Section IV. presents the
transformation of the Hamiltonian in a positive semidefinite form, Section V.
solves the matching conditions, Section VI. presents the deduction strategy
for the exact ground states, and the obtained ground states 
together with their physical properties, and 
Section VII. containing the summary and conclusions 
closes the presentation. The Appendix contains mathematical details
related to the exemplification in a simple case of the deduced ground states.

\section{The system of interest}

The zig-zag hexagon chain is presented in Fig.1. In the cell constructed at 
the site ${\bf i}$ one has 4 lattice sites whose position relative to the
site ${\bf i}$ is given by the vectors ${\bf r}_{\nu}$, $\nu=1,2,3,4$. The
system contains four sublattices $S_{\nu}$ containing the lattice sites
${\bf i}+{\bf r}_{\nu} \in S_{\nu}$, hence $\nu$ represents as well the
sublattice index. For mathematical simplicity one considers ${\bf r}_1=0$
during the calculations.
The primitive Bravais vector ${\bf a}$ is directed along the
line of the chain. The hexagon side is $b=a/\sqrt{3}$, where $a=|{\bf a}|$ is
the lattice constant.
One considers below a chain build up from $N_c$ cells,
the number of lattice sites being $N_{\Lambda}=4N_c$, while the number of
electrons is denoted by $N$. 

The Hamiltonian of the system $\hat H = \hat H_0 + \hat H_U$ is given by
\begin{eqnarray}
\hat H_0 &=&
\sum_{\sigma} \sum_{{\bf i}=1}^{N_c} \{ \: [ t ( \hat c^{\dagger}_{{\bf i}+
{\bf r}_2,\sigma} \hat c_{{\bf i}+{\bf r}_1,\sigma} +  \hat c^{\dagger}_{
{\bf i}+ {\bf r}_1 + {\bf a},\sigma} \hat c_{{\bf i}+{\bf r}_2,\sigma} +
\hat c^{\dagger}_{{\bf i}+ {\bf r}_4,\sigma} \hat c_{{\bf i}+{\bf r}_3 +
{\bf a},\sigma} + \hat c^{\dagger}_{{\bf i}+ {\bf r}_3,\sigma} 
\hat c_{{\bf i}+{\bf r}_4,\sigma}) 
\nonumber\\
&+& t_1 \hat c^{\dagger}_{{\bf i}+ {\bf r}_1,
\sigma} \hat c_{{\bf i}+{\bf r}_3,\sigma} + H.c. ] +
[ t' ( \hat c^{\dagger}_{{\bf i}+ {\bf r}_2,\sigma} \hat c_{{\bf i}+{\bf r}_3,
\sigma} + \hat c^{\dagger}_{{\bf i}+ {\bf r}_3 + {\bf a},\sigma} 
\hat c_{{\bf i}+{\bf r}_2,\sigma} + \hat c^{\dagger}_{{\bf i}+ {\bf r}_4,
\sigma} \hat c_{{\bf i}+{\bf r}_1+{\bf a},\sigma} 
\nonumber\\
&+& \hat c^{\dagger}_{{\bf i}+
{\bf r}_1,\sigma} \hat c_{{\bf i}+{\bf r}_4,\sigma} ) + t'_1 (
\hat c^{\dagger}_{{\bf i}+{\bf r}_1+{\bf a},\sigma} \hat c_{{\bf i}+{\bf r}_1,
\sigma}  + \hat c^{\dagger}_{{\bf i}+{\bf r}_3+{\bf a},\sigma} \hat c_{{\bf i}
+{\bf r}_3,\sigma} ) 
\nonumber\\
&+& 
t_e(\hat c^{\dagger}_{{\bf i}+{\bf r}_2+{\bf a},\sigma} \hat c_{{\bf i}+
{\bf r}_2,\sigma} + \hat c^{\dagger}_{{\bf i}+{\bf r}_4+{\bf a},\sigma} 
\hat c_{{\bf i}+{\bf r}_4,\sigma})  + H.c. ] 
\nonumber\\
&+& [ \epsilon_0 ( \hat n_{{\bf i}+{\bf r}_2,
\sigma} + \hat n_{{\bf i}+{\bf r}_4,\sigma} ) + \epsilon_1 ( 
\hat n_{{\bf i}+{\bf r}_1,\sigma} + \hat n_{{\bf i}+{\bf r}_3,\sigma} ) ] ,
\nonumber\\
\hat H_U &=& U \sum_{{\bf i}=1}^{N_c} \sum_{\nu=1}^4 \hat n_{{\bf i}+{\bf r}_{
\nu},\uparrow} \hat n_{{\bf i}+{\bf r}_{\nu},\downarrow} ,
\label{e1}
\end{eqnarray}
where $\hat c^{\dagger}_{{\bf j},\sigma}$ is the canonical Fermi operator 
creating an electron at the site ${\bf j}$ with spin projection $\sigma$,
and
$\hat n_{{\bf i},\sigma}=\hat c^{\dagger}_{{\bf i},\sigma} \hat c_{{\bf i},
\sigma}$ represents
the particle number operator for electrons with spin $\sigma$
placed at the lattice site ${\bf i}$. Since real systems are in view, several
hopping matrix elements are considered in (\ref{e1}). The
 $t$ and $t_1$ terms characterize nearest neighbor 
hopping matrix elements ($t_1$ is placed perpendicular to the line of the 
chain). The $t', \: t'_1, \: t_e$ contributions represent next nearest 
neighbor hopping matrix 
elements ($t'_1$ and $t_e$ are directed along the line of the chain, $t'_1$ 
connects internal sites while $t_e$ external sites of the chain). 
The $\epsilon_0$, $\epsilon_1$ parameters are on-site one-particle potentials 
($\epsilon_0$ is placed on external sites), see Fig.2. 
Finally, $\hat H_U$ characterizes the 
on-site Coulomb repulsion where $U>0$ is considered.

\begin{figure}[t!]
\epsfxsize=8cm
\centerline{\epsfbox{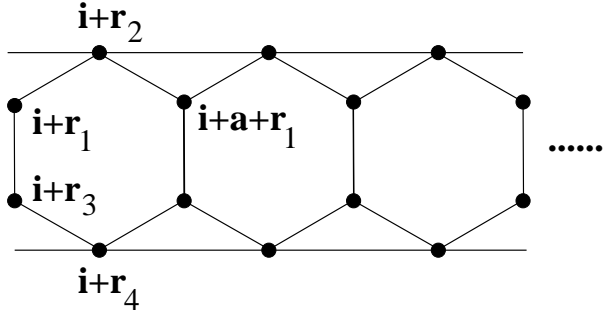}}
\caption{The studied zig-zag hexagon chain. The cell constructed
at the site ${\bf i}$ contains 4 sites whose in-cell positions relative to
${\bf i}$ are specified by the vectors ${\bf r}_{\nu}$, $\nu=1,2,3,4$,
where, for convenience one takes ${\bf r}_1=0$. The vector ${\bf a}$ 
represents the unique primitive vector of the Bravais lattice. In the same 
time, ${\bf r}_{\nu}$ denotes lattice sites in four different sublattices
$S_{\nu}$.}
\label{fig1}
\end{figure}

\begin{figure}[t!]
\epsfxsize=8cm
\centerline{\epsfbox{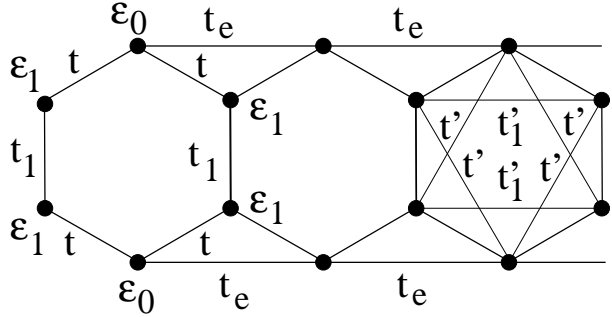}}
\caption{The parameters of the kinetic part of the Hamiltonian. The
$t$ and $t_1$ terms
represent nearest neighbor hopping matrix elements
from which $t_1$ is placed
along the touching bonds between hexagons. $t'$, $t'_1$ and $t_e$ are hopping
matrix elements describing next nearest neighbor hoppings. From these $t_e$ 
describes the external
hopping relative to hexagons, while $t'_1$ is parallel to the line 
of the chain, being placed inside the hexagons. Finally, $\epsilon_1$, 
($\epsilon_0$) represents the on-site potential
on contact points between hexagons (external sites of hexagons). 
For the clarity 
of the notations, the next nearest neighbor hoppings are separately presented
in the third cell of the figure.}
\label{fig2}
\end{figure}

\section{The non-interacting band structure}

\subsection{The deduction of the bare band structure}

In order to deduce the non-interacting band structure one transforms $\hat H_0$
from (\ref{e1}) in ${\bf k}$-space. For this one Fourier transforms the
$\hat c_{{\bf j},\sigma}$ operators as follows
\begin{eqnarray}
\hat c_{{\bf i}+{\bf r}_{\nu}+{\bf r},\sigma}=\frac{1}{\sqrt{N_c}}
\sum_{{\bf k}=1}^{N_c} \hat c_{\nu,{\bf k},\sigma} e^{-i {\bf k}({\bf i}+
{\bf r}_{\nu}+{\bf r})} ,
\label{f1}
\end{eqnarray}
where $\hat c_{\nu,{\bf k},\sigma}$ represents the annihilation operator
for the state $({\bf k},\sigma)$ in the sublattice $\nu$. 
Note that  one has 
${\bf r}_1=0$, and ${\bf r}$ from (\ref{f1}) takes two possible 
values, namely $0$ and ${\bf a}$. Substituting (\ref{f1}) in (\ref{e1}),
one finds
\begin{eqnarray}
\hat H_0 &=& \sum_{\sigma} \sum_{\bf k} \{ [ t_{1,2}({\bf k}) 
\hat c^{\dagger}_{
1,{\bf k},\sigma} \hat c_{2,{\bf k},\sigma} + t_{1,3}({\bf k}) 
\hat c^{\dagger}_{1,{\bf k},\sigma} \hat c_{3,{\bf k},\sigma} + 
t_{3,4}({\bf k}) \hat c^{\dagger}_{3,{\bf k},\sigma} \hat c_{4,{\bf k},\sigma}
+ H.c.] 
\nonumber\\
&+& [t_{1,4}({\bf k}) \hat c^{\dagger}_{1,{\bf k},\sigma} \hat c_{4,
{\bf k},\sigma} + t_{2,3}({\bf k}) \hat c^{\dagger}_{2,{\bf k},\sigma} 
\hat c_{3,{\bf k},\sigma} + t_{1,1}({\bf k}) \hat c^{\dagger}_{1,{\bf k},
\sigma} \hat c_{1,{\bf k},\sigma} 
\nonumber\\
&+& t_{2,2}({\bf k}) \hat c^{\dagger}_{2,{\bf k},\sigma} 
\hat c_{2,{\bf k},\sigma}
+  t_{3,3}({\bf k}) \hat c^{\dagger}_{3,{\bf k},\sigma} \hat c_{3,{\bf k},
\sigma} +  t_{4,4}({\bf k}) \hat c^{\dagger}_{4,{\bf k},\sigma} 
\hat c_{4,{\bf k},
\sigma} + H.c.] 
\nonumber\\
&+&
[\epsilon_0(\hat c^{\dagger}_{2,{\bf k},\sigma} \hat c_{2,{\bf k},\sigma} +
\hat c^{\dagger}_{4,{\bf k},\sigma} \hat c_{4,{\bf k},\sigma}) +
\epsilon_1(\hat c^{\dagger}_{1,{\bf k},\sigma} \hat c_{1,{\bf k},\sigma} +
\hat c^{\dagger}_{3,{\bf k},\sigma} \hat c_{3,{\bf k},\sigma}) ] \} ,
\label{f2}
\end{eqnarray}
where the first, second and third, and fourth rows represent in order the 
nearest neighbor,
next nearest neighbor, and on-site contributions. Furthermore one has
\begin{eqnarray}
&&t_{1,2}({\bf k})= t( e^{+i{\bf k}({\bf a}-{\bf r}_2)}+e^{-i{\bf k}{\bf r}_2}
), \quad t_{1,3}({\bf k})= t_1 e^{-i{\bf k}{\bf r}_3}, \quad
t_{3,4}({\bf k}) = t(e^{+i{\bf k}({\bf r}_3-{\bf r}_4)} +
e^{-i{\bf k}({\bf r}_4-{\bf r}_3-{\bf a})}),
\nonumber\\
&&t_{1,4}({\bf k})=t'(e^{-i{\bf k}{\bf r}_4} +e^{-i{\bf k}({\bf r}_4-{\bf a})}
), \quad t_{2,3}({\bf k})=t'(e^{+i{\bf k}({\bf r}_2-{\bf r}_3)}+e^{-i{\bf k}(
{\bf r}_3+{\bf a}-{\bf r}_2)}), 
\nonumber\\
&&t_{1,1}({\bf k})= t'_1 e^{+i{\bf k}{\bf a}} , \quad t_{2,2}({\bf k})= t_e
e^{+i{\bf k}{\bf a}}, \quad t_{3,3}({\bf k}) = t'_1 e^{+i{\bf k}{\bf a}},
\quad t_{4,4}({\bf k}) = t_e e^{+i{\bf k}{\bf a}} ,
\label{f3}
\end{eqnarray}
where for ${\bf r}_{\nu}$, $\nu=2,3,4$, see Fig.1, while
${\bf a}$ represents the primitive vector of the Bravais lattice. Taking into 
consideration a Cartesian system of coordinates with versors $({\bf i}_1,
{\bf j}_1)$ whose $x$-axis is directed along the line of the chain, one has
\begin{eqnarray}
{\bf r}_1=0, \quad {\bf r}_2= \frac{a}{2} {\bf i}_1 + \frac{a}{2\sqrt{3}}
{\bf j}_1, \quad {\bf r}_3 = - \frac{a}{\sqrt{3}} {\bf j}_1, \quad {\bf r}_4=
\frac{a}{2}{\bf i}_1 - \frac{a\sqrt{3}}{2} {\bf j}_1,
\label{f4}
\end{eqnarray}
where $a=|{\bf a}|$ holds.

Introducing the $1\times 4$ row vector ${\bf C}^{\dagger}$ and its $4\times 1$
adjoint ${\bf C}$ by
\begin{eqnarray}
{\bf C}^{\dagger} = ( \hat c^{\dagger}_{1,{\bf k},\sigma}, 
\hat c^{\dagger}_{2,{\bf k},\sigma}, \hat c^{\dagger}_{3,{\bf k},\sigma},
\hat c^{\dagger}_{4,{\bf k},\sigma}), \quad
\quad
{\bf C} =
\left( \begin{array}{c}
\hat c_{1,{\bf k},\sigma}\\
\hat c_{2,{\bf k},\sigma}\\
\hat c_{3,{\bf k},\sigma}\\
\hat c_{4,{\bf k},\sigma}\\ 
\end{array} \right) ,
\label{f5}
\end{eqnarray}
one observes that (\ref{f2}) can be written in a matrix product form, namely
\begin{eqnarray}
\hat H_0 = \sum_{\sigma} \sum_{\bf k} {\bf C}^{\dagger} \tilde {\bf M} 
{\bf C} ,
\label{f6}
\end{eqnarray}
where, introducing the notation $f({\bf k}) = (1+e^{+i{\bf k}{\bf a}})$,
for the matrix $\tilde {\bf M}$ one has
\begin{eqnarray}
\tilde {\bf M} =
\left( \begin{array}{cccc}
\epsilon_1 + 2t'_1 \cos {\bf a}{\bf k} & t e^{-i{\bf k}{\bf r}_2} f({\bf k}) 
& t_1 e^{-i{\bf k}{\bf r}_3} & t' e^{-i{\bf k}{\bf r}_4} f({\bf k}) \\
 t e^{+i{\bf k}{\bf r}_2} f^*({\bf k})   & \epsilon_0  +2t_e \cos{\bf a}{\bf k}
& t'e^{-i{\bf k}({\bf r}_3-{\bf r}_2)} f^*({\bf k})  & 0 \\
t_1 e^{+i{\bf k}{\bf r}_3} & t'e^{+i{\bf k}({\bf r}_3-{\bf r}_2)} f({\bf k})
& \epsilon_1 + 2t'_1 \cos {\bf a}{\bf k} & te^{-i{\bf k}({\bf r}_4-{\bf r}_3)}
f({\bf k}) \\
t'e^{+i{\bf k}{\bf r}_4} f^*({\bf k}) & 0 & te^{+i{\bf k}({\bf r}_4-{\bf r}_3)}
f^*({\bf k})    & \epsilon_0 +2 t_e \cos {\bf a}{\bf k} \\
\end{array} \right) .
\label{f7}
\end{eqnarray} 
The bare band structure is obtained from the secular equation of the matrix 
$\tilde {\bf M}$ from (\ref{f7}) which provides
\begin{eqnarray}
&&(\epsilon_0-\lambda + 2t_e \cos{\bf a}{\bf k} )^2 [(\epsilon_1-\lambda
+2t'_1\cos{\bf a}{\bf k})^2 -t^2_1] +
(\epsilon_0-\lambda + 2t_e \cos {\bf a}{\bf k}) |f({\bf k})|^2 
\nonumber\\
&&\times [2tt_1t'- 
(\epsilon_1-\lambda +2t'_1 \cos {\bf a}{\bf k})({t'}^2+t^2)] 
-t |f({\bf k})|^2 [t(\epsilon_0-\lambda +2t_e \cos{\bf a}{\bf k})(
\epsilon_1-\lambda +2t'_1\cos{\bf a}{\bf k})
\nonumber\\
&&-(\epsilon_0-\lambda + 2t_e \cos {\bf a}{\bf k})t't_1 + 
|f({\bf k})|^2 t ({t'}^2-t^2)]
-t' |f({\bf k})|^2 [t'(\epsilon_0-\lambda +2t_e \cos{\bf a}{\bf k})
\nonumber\\
&&\times (\epsilon_1-\lambda +2t'_1\cos{\bf a}{\bf k})-
(\epsilon_0-\lambda+ 2t_e \cos{\bf a}{\bf k} )tt_1 + 
|f({\bf k})|^2 t' (t^2-{t'}^2)] =0 .
\label{f8}
\end{eqnarray}
The non-interacting band structure containing 4 bands is obtained from the
solutions $E_{\eta}({\bf k}) = \lambda$ of the quadratic algebraic equation
(\ref{f8}), which provides four solutions $\eta=1,2,3,4$. One further notes 
that in (\ref{f8}), $|f({\bf k})|^2=2(1+\cos {\bf a}{\bf k})$ holds, and 
$k={\bf a}{\bf k} \in (-\pi, \pi]$ is satisfied for the first Brillouin zone.

Introducing the notation $\bar \epsilon_{\alpha}=\epsilon_{\alpha}-\lambda$
for $\alpha=0,1$, from (\ref{f8}) one obtains the equation for the bare
band structure in the form
\begin{eqnarray}
&&(\bar \epsilon_0 + 2t_e \cos k)^2 [(\bar \epsilon_1 + 2 t'_1 \cos k)^2
-t_1^2] - 4 (\bar \epsilon_0 + 2t_e \cos k) (1+\cos k)
\nonumber\\
&&\times [(\bar \epsilon_1 + 
2 t'_1 \cos k)(t^2+{t'}^2) - 2tt_1t'] 
+ 4(1+\cos k)^2 ({t'}^2-t^2)^2 =0 .
\label{f9}
\end{eqnarray}

\begin{figure}[t!]
\epsfxsize=9cm
\centerline{\epsfbox{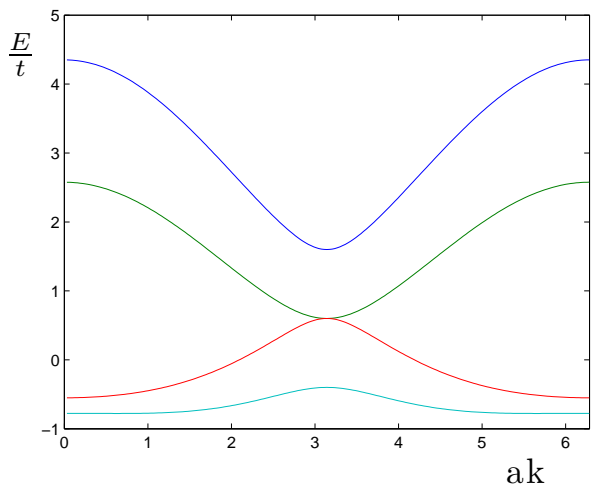}}
\caption{An exemplifying image for the band structure plotted on a 
$2\pi$ domain for
the $k$ variable. The used parameters are $t_1/t=1, 
t'/t=0.2, t'_1/t=0.2, t_e/t=0.2, \epsilon_0/t=1, \epsilon_1/t=1$. 
By modifying the $\hat H_0$ parameter values and signs, the relative 
inter-band distances and band shapes can be changed.}
\label{fig3}
\end{figure}
\begin{figure}[t!]
\epsfxsize=9cm
\centerline{\epsfbox{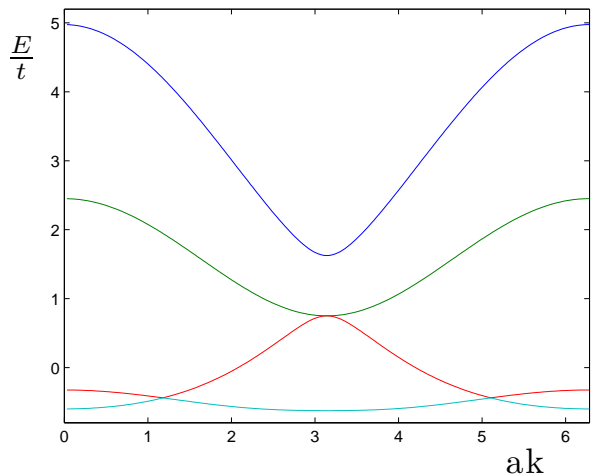}}
\caption{Band structure with cusp points plotted at $t_1/t=1.8/1.6, 
t'/t=0.5/1.6, t'_1/t=0.6/1.6, t_e/t=0.2/1.6, \epsilon_0/t=1, 
\epsilon_1/t=2/1.6$ parameter values.}
\label{fig4}
\end{figure}
\begin{figure}[t!]
\epsfxsize=9cm
\centerline{\epsfbox{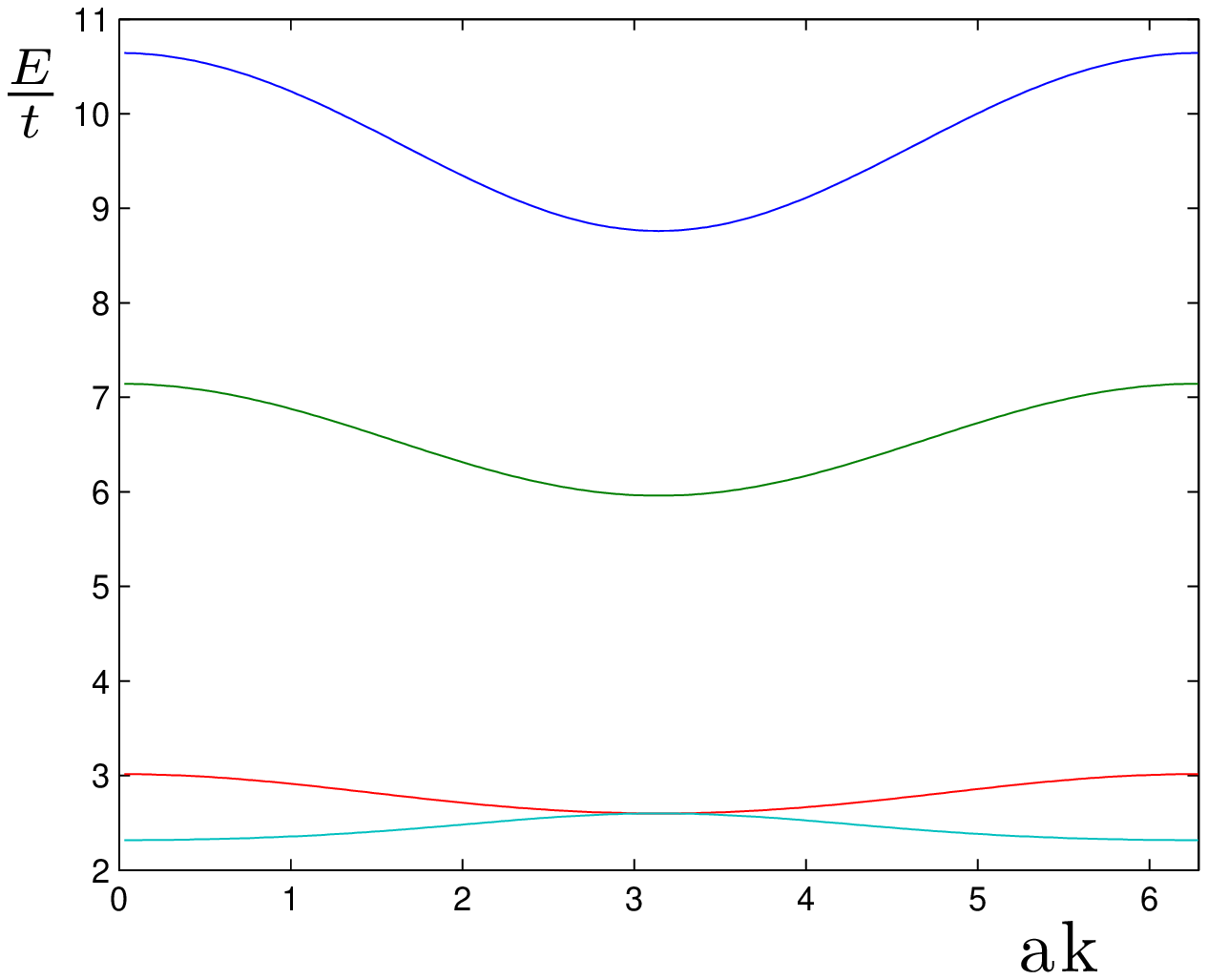}}
\caption{Band structure with closely situated intersection points 
plotted at $t_1/t=1.4, 
t'/t=0.2, t'_1/t=0.2, t_e/t=0.2, \epsilon_0/t=3, 
\epsilon_1/t=7.76$ parameter values.}
\label{fig5}
\end{figure}

\subsection{The properties of the bare band structure}

A picture representing an exemplification for the bare band structure 
is presented in 
Fig.3.  Modifying the parameter values, one obtains cusp points which can be
considered in a given extent reminiscent
of the Dirac points present in the graphene case \cite{pr}, see Fig.4. 
But since in
the studied case, the $|k\pm k^*|$ values not provide the same energy 
around the cusp points $k^*$, for small $k$, 
Dirac points in rigorous terms are not present here. 
Another case with closely situated intersection 
points for the lower two bands is presented in Fig.5.

It is important to underline that flat bands are not possible to occur for
the studied system. In order to show this one writes (\ref{f9}) into the form
\begin{eqnarray}
A_4 \cos^4 k + A_3 \cos^3 k + A_2 \cos^2 k + A_1 \cos k + A_0 =0,
\label{f10}
\end{eqnarray}
where $A_n=A_n(\epsilon_0,\epsilon_1,t,t',t_1,t'_1,t_e,\lambda)$ for all
$n=0,1,...,4$
are $k$ independent. Flat bands (e.g. $k$ independences in $\lambda$) are 
obtained if simultaneously $A_n=0$ holds for all $n$. This provides
the following system of equations
\begin{eqnarray}
&&A_4(\epsilon_0,\epsilon_1,t,t_1,t',t_1',t_e,\lambda) = 16 t_e^2 t_1'^2
\nonumber \\
&&A_3(\epsilon_0,\epsilon_1,t,t_1,t',t_1',t_e,\lambda) = 16 (\epsilon_0-
\lambda)t_e t_1'^2 +16(\epsilon_1-\lambda) t_e^2 t_1' - 16t_et_1'(t^2+t'^2)
\nonumber \\
&&A_2(\epsilon_0,\epsilon_1,t,t_1,t',t_1',t_e,\lambda) = 4 t_1'^2(\epsilon_0-
\lambda)^2  +16 (\epsilon_0-\lambda)(\epsilon_1-\lambda)t_e t_1'   
+ 4t_e^2(\epsilon_1-\lambda)^2   -4 t_e^2t_1^2
\nonumber \\
&&\;\;\;\;\;\;\;\;\;-16t_et_1'(t^2+t'^2) -8t_1'(\epsilon_0-\lambda) (t^2+t'^2)
-8 t_e(\epsilon_1-\lambda) (t^2+t'^2) + 16 t t_1 t' t_e
+4 (t'^2-t^2)^2
\nonumber \\
&&A_1(\epsilon_0,\epsilon_1,t,t_1,t',t_1',t_e,\lambda) = 4(\epsilon_0-
\lambda)^2(\epsilon_1-\lambda)t_1'  + 4(\epsilon_0-\lambda)(\epsilon_1-
\lambda)^2 t_e  -4 t_1^2 t_e(\epsilon_0-\lambda)
\nonumber \\
&&\;\;\;\;\;\;\;\;\;
 - 8 t_1'(\epsilon_0-\lambda) (t^2+t'^2) - 8t_e(\epsilon_1-\lambda) (t^2+t'^2)
-4(\epsilon_0-\lambda) (\epsilon_1-\lambda) (t^2+t'^2)
\nonumber \\
&&\;\;\;\;\;\;\;\;\;
+ 8t t_1 t' (\epsilon_0-\lambda)
+ 16 t t_1 t' t_e  + 8 (t'^2-t^2)^2
\nonumber \\
&&A_0(\epsilon_0,\epsilon_1,t,t_1,t',t_1',t_e,\lambda) = 
(\epsilon_0-\lambda)^2(\epsilon_1-\lambda)^2 -t_1^2(\epsilon_0-\lambda)^2  
-4(\epsilon_0-\lambda)(\epsilon_1-\lambda)(t^2+t'^2)
\nonumber \\
&&\;\;\;\;\;\;\;\;\;+8t t_1 t' (\epsilon_0-\lambda)+4(t'^2-t^2)^2 .
\label{f11}
\end{eqnarray}
As seen, (\ref{f11}) presents $A_n=0$ solutions for all $n$ only at 
$t_e t'_1 =0$, hence taking
non-zero values for all introduced hopping terms, flat bands are not possible 
to occur into the system.

\section{The transformation of the Hamiltonian in positive semidefinite form}

One introduces four block operators $\hat A_{\nu,{\bf i},\sigma}$,
$\nu=1,2,3,4$ defined in the cell constructed at the site ${\bf i}$
by the following relations
\begin{eqnarray}
&&\hat A_{1,{\bf i},\sigma}= a_1 \hat c_{{\bf i}+{\bf r}_1,\sigma} +
a_2 \hat c_{{\bf i}+{\bf r}_2,\sigma} +a_3 \hat c_{{\bf i}+{\bf r}_3,\sigma} +
a_0 \hat c_{{\bf i}-{\bf a}+{\bf r}_2,\sigma},
\nonumber\\
&&\hat A_{2,{\bf i},\sigma}= b_1 \hat c_{{\bf i}+{\bf r}_1,\sigma} +
b_3 \hat c_{{\bf i}+{\bf r}_3,\sigma} +b_4 \hat c_{{\bf i}+{\bf r}_4,\sigma} +
b_0 \hat c_{{\bf i}-{\bf a}+{\bf r}_4,\sigma},
\nonumber\\
&&\hat A_{3,{\bf i},\sigma}= d_2 \hat c_{{\bf i}+{\bf r}_2,\sigma} +
d_3 \hat c_{{\bf i}+{\bf r}_3,\sigma} + d_0 \hat c_{{\bf i}+{\bf a}+{\bf r}_3,
\sigma},
\nonumber\\
&&\hat A_{4,{\bf i},\sigma}= e_1 \hat c_{{\bf i}+{\bf r}_1,\sigma} +
e_4 \hat c_{{\bf i}+{\bf r}_4,\sigma} + e_0 \hat c_{{\bf i}+{\bf a}+{\bf r}_1,
\sigma},
\label{e2}
\end{eqnarray}
The lattice sites on which the operators $\hat A_{\nu,{\bf i},\sigma}$ are 
defined for the cell constructed at the site ${\bf i}$ are presented in Fig.6.

\begin{figure}[t!]
\epsfxsize=8cm
\centerline{\epsfbox{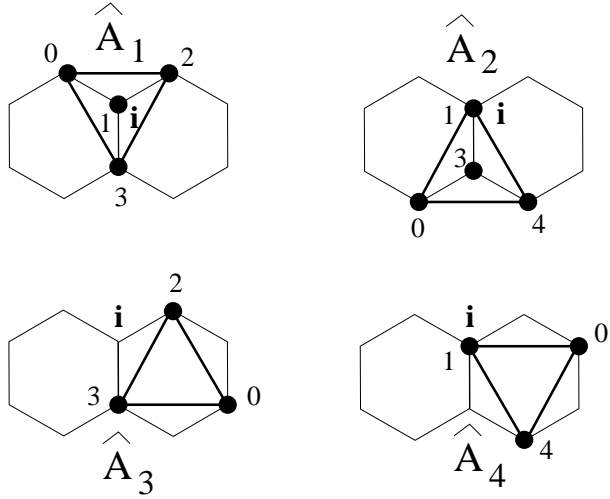}}
\caption{The lattice sites on which the block operators $\hat A_{\nu,{\bf i},
\sigma}$, $\nu=1,2,3,4$ are defined for the cell constructed at the site 
${\bf i}$. In each figure the position of the site ${\bf i}$ is explicitly
shown. The numbers are representing the numbering of the coefficients of
$\hat A_{\nu,{\bf i},\sigma}$ operators (see also (\ref{e2})).}
\label{fig6}
\end{figure}

Using periodic boundary conditions and based on (\ref{e2}), the Hamiltonian of 
the problem transforms into a positive semidefinite form as follows

\begin{eqnarray}
\hat H = \sum_{{\bf i}=1}^{N_c} \sum_{\sigma}\sum_{\nu=1}^{4}
\hat A^{\dagger}_{\nu,{\bf i},\sigma} \hat A_{\nu,{\bf i},\sigma} +
\hat H_U + K \hat N ,
\label{e3}
\end{eqnarray}
where $\hat N$ denotes the operator of the 
total number of particles. 
The relation (\ref{e3}) holds if the following
matching conditions are satisfied
\begin{eqnarray}
&& t = a^*_2 a_1 = a^*_1 a_0 = b^*_4 b_3 = b^*_3 b_0 ,
\nonumber\\
&&t_e = a^*_2 a_0 = b^*_4 b_0 ,
\nonumber\\
&&t_1 = a^*_3 a_1 + b^*_3 b_1 ,
\nonumber\\
&& t' = a^*_2 a_3 + d_2^* d_3  = a^*_3 a_0 + d^*_0 d_2 =
b^*_4 b_1 + e^*_4 e_1 = b^*_1 b_0 + e^*_0 e_4 ,
\nonumber\\
&& t'_1 = e^*_0 e_1 = d_0^* d_3 ,
\nonumber\\
&&\epsilon_0 - K = |a_0|^2 + |a_2|^2 + |d_2|^2 = |b_0|^2 + |b_4|^2 + |e_4|^2 ,
\nonumber\\
&&\epsilon_1 - K = |a_1|^2 + |e_0|^2 + |e_1|^2 + |b_1|^2 =
|b_3|^2 + |d_0|^2 +|d_3|^2 + |a_3|^2 .
\label{e4}
\end{eqnarray}
The matching conditions (\ref{e4}) have been obtained by calculating
$\sum_{{\bf i}=1}^{N_c} \sum_{\sigma}\sum_{\nu=1}^{4}
\hat A^{\dagger}_{\nu,{\bf i},\sigma} \hat A_{\nu,{\bf i},\sigma}$ from
(\ref{e3}) and equating the result term by term to the expression of $\hat H_0$
from (\ref{e1}).

\section{Deduction of $\hat A_{\nu,{\bf i},\sigma}$ operators}

In deducing the block operators defined in (\ref{e2}) one must calculate the
prefactors $a_{\alpha},b_{\alpha},d_{\alpha},e_{\alpha}$ present in equations
(\ref{e2}). These are obtained by solving the matching conditions (\ref{e4}).

Solving the equations present in (\ref{e4}) one proceeds as follows.
The first two rows of Eq.(\ref{e4}) provide solutions only for $t_e >0$
and taking into account real hopping matrix elements, give 
\begin{eqnarray}
&&|a_1|=|b_3|=\frac{t}{\sqrt{t_e}},
\nonumber\\
&&a_2 = a_0 = \frac{t}{a_1^*}, \quad b_4 = b_0 = \frac{t}{b^*_3} .
\label{e5}
\end{eqnarray}
The fifth row of (\ref{e4}) provides
\begin{eqnarray}
e_0=\frac{t'_1}{e^*_1}, \quad d_0=\frac{t'_1}{d^*_3} ,
\label{e6}
\end{eqnarray}
and the remaining equations from (\ref{e4}) transform into the form
\begin{eqnarray}
&&t_1=a^*_3a_1 + a_1^* b_1 e^{-i\phi^a_1},
\nonumber\\
&&t'=t\frac{a_3}{a_1} +d_2^* d_3= t \frac{a_3}{a_1} + t'_1 \frac{d^*_2}{d_3^*},
\nonumber\\
&&t'=t\frac{b_1}{b_3} +e_4^* e_1= t \frac{b_1}{b_3} + t'_1 \frac{e^*_4}{e_1^*},
\nonumber\\
&&\epsilon_0-K = 2t_e+|d_2|^2 = 2t_e +|e_4|^2,
\nonumber\\
&&\epsilon_1-K=\frac{t^2}{t_e} + \frac{{t'_1}^2}{|e_1|^2} + |e_1|^2 + |b_1|^2 =
\frac{t^2}{t_e} + \frac{{t'_1}^2}{|d_3|^2} + |d_3|^2 + |a_3|^2.
\label{e7}
\end{eqnarray}
The second and third line of (\ref{e7}) arises from the fourth row of
(\ref{e4}).  For the first line of (\ref{e7}), given by $|a_1|=|b_3|$ in 
(\ref{e5}), one uses 
$b_3=a_1e^{i\phi^a_1}$. Now from the second and third line of (\ref{e7}), 
besides the condition $t'_1>0$, one finds $|d_3|=|e_1|=\sqrt{t'_1}$. 
Consequently, the last two equations from (\ref{e7}) give $|d_2|=|e_4|$,
and $|b_1|=|a_3|$. Using $b_1=a_3 e^{i\phi^a_3}$ from the previous 
condition, the first equation of (\ref{e7}) becomes
\begin{eqnarray}
t_1=a^*_3a_1 + a_1^* a_3 e^{i(\phi^a_3-\phi^a_1)},
\label{e8}
\end{eqnarray}
which, because of the real nature of $t_1$ gives $\phi^a_3=\phi^a_1$, from 
where
$t_1=a^*_3a_1+a^*_1a_3=2Re(a^*_1a_3)$, furthermore $b_1/b_3=a_3/a_1$
arises. At this step one observes that
in (\ref{e2}) each $\hat A_{\nu,{\bf i},\sigma}$ operator can be multiplied
by an arbitrary phase factor without changing the expression of the
transformed Hamiltonian (\ref{e3}). Consequently, by this multiplication
relating $\hat A_{1,{\bf i},\sigma}$, the prefactor
$a_1$ can be taken real,
hence from the expression of $t_1$ the parameter $a_3$ must also be real.
As a consequence, $t_1=2a_1 a_3$ holds, from where 
\begin{eqnarray}
a_3=\frac{t_1}{2a_1}= \frac{t_1\sqrt{t_e}}{2t}, \quad a_1=\frac{t}{\sqrt{t_e}}.
\label{e9}
\end{eqnarray}
As a result, one finds that the coefficients
$a_{\alpha}$ are given by
\begin{eqnarray}
a_0=\sqrt{t_e}, \quad a_1=\frac{t}{\sqrt{t_e}}, \quad a_2=\sqrt{t_e}, \quad
a_3=\frac{t_1\sqrt{t_e}}{2t}.
\label{e10}
\end{eqnarray}

In what follows one concentrates on the $b_{\alpha}$ coefficients. Given by
$b_1=a_3 e^{i\phi^a_3}, b_3=a_1e^{i\phi^a_1}$ where $a_1,a_3$ are real and
$\phi^a_1=\phi^a_3$, a multiplication by $e^{-i\phi^a_1}$ of 
$\hat A_{2,{\bf i},\sigma}$ from (\ref{e2}) provides real $b_1$ and $b_3$.
Hence (\ref{e5},\ref{e10}) together with $b_1/b_3=a_3/a_1=t_1t_e/(2t^2)$ gives
\begin{eqnarray}
b_0=\sqrt{t_e}, \quad b_1=\frac{t_1\sqrt{t_e}}{2t}, \quad b_3=\frac{t}{
\sqrt{t_e}}, \quad b_4=\sqrt{t_e} .
\label{e11}
\end{eqnarray}

At this moment one continues the calculation by deducing $e_{\alpha}$ and
$d_{\alpha}$. From the real value of $t'$, taking into account the second and
third expression of (\ref{e7}), one observes that by fixing $d_3$ 
(similarly $e_1$) to be real by multiplication
 by a phase factor, then $d_2$ 
(similarly $e_4$) also becomes real. Consequently $t'=(t_1t_e)/(2t)+d_2
\sqrt{t'_1} = (t_1t_e)/(2t)+e_4\sqrt{t'_1}$ and $d_3=e_1=\sqrt{t'_1}$ together
with Eq.(\ref{e6}) provides
\begin{eqnarray}
&&e_0 = \sqrt{t'_1}, \quad e_1 = \sqrt{t'_1}, \quad e_4=\frac{2 t t'-t_1t_e}{
2t \sqrt{t'_1}},
\nonumber\\
&&d_0= \sqrt{t'_1}, \quad d_2=\frac{2 t t'-t_1t_e}{2t \sqrt{t'_1}}, \quad
d_3=\sqrt{t'_1}.
\label{e12}
\end{eqnarray}
From the last two lines of (\ref{e7}) one further finds the $K$ value
\begin{eqnarray}
K=\epsilon_0-2t_e - \frac{(2tt'-t_1t_e)^2}{4t^2t'_1},
\label{e13}
\end{eqnarray}
and a condition which must be satisfied by the Hamiltonian parameters, namely
\begin{eqnarray}
\epsilon_0=\epsilon_1+2(t_e-t'_1)- \frac{t^2}{t_e} - \frac{t_1^2t_e}{4t^2}+
\frac{(2tt'-t_1t_e)^2}{4t^2t'_1}.
\label{e14}
\end{eqnarray}
The equality (\ref{e14}) represents the requirement for the on-site potential
$\epsilon_0$ necessary to be satisfied in order to find the expression for the transformed 
Hamiltonian (\ref{e3}) a valid relation. Besides
(\ref{e14}), the parameter space region 
where the transformation from (\ref{e1})
to the positive semidefinite form (\ref{e3}) of the Hamiltonian is valid, is
described by $t_e>0, t'_1 >0$. In the presence of all these conditions,
the bare band structure is exemplified by Fig.5.

Using the results presented in (\ref{e10}-\ref{e12}) one finds for the block
operators introduced in (\ref{e2}) the explicit expressions
\begin{eqnarray}
&&\hat A_{1,{\bf i},\sigma}= \sqrt{t_e} (\hat c_{{\bf i}+{\bf r}_2-{\bf a},
\sigma} + \hat c_{{\bf i}+{\bf r}_2,\sigma}) +\frac{t}{\sqrt{t_e}} 
\hat c_{{\bf i}+{\bf r}_1,\sigma} + \frac{t_1\sqrt{t_e}}{2t}
\hat c_{{\bf i}+{\bf r}_3,\sigma},
\nonumber\\
&&\hat A_{2,{\bf i},\sigma}= \sqrt{t_e} (\hat c_{{\bf i}+{\bf r}_4-{\bf a},
\sigma} + \hat c_{{\bf i}+{\bf r}_4,\sigma}) + \frac{t}{\sqrt{t_e}}  
\hat c_{{\bf i}+{\bf r}_3,\sigma} + \frac{t_1\sqrt{t_e}}{2t}
\hat c_{{\bf i}+{\bf r}_1,\sigma},
\nonumber\\
&&\hat A_{3,{\bf i},\sigma}= \sqrt{t'_1}(\hat c_{{\bf i}+{\bf r}_3,\sigma} +
\hat c_{{\bf i}+{\bf r}_3+{\bf a},\sigma}) + \frac{2tt'-t_1t_e}{2t\sqrt{t'_1}} 
\hat c_{{\bf i}+{\bf r}_2,\sigma},
\nonumber\\
&&\hat A_{4,{\bf i},\sigma}= \sqrt{t'_1} (\hat c_{{\bf i}+{\bf r}_1,\sigma} +
\hat c_{{\bf i}+{\bf a}+{\bf r}_1,\sigma}) + \frac{2tt'-t_1t_e}{
2t\sqrt{t'_1}} \hat c_{{\bf i}+{\bf r}_4,\sigma}.
\label{e15}
\end{eqnarray}

\section{The deduction of exact ground states}

The ground states are deduced in two steps. i) First one looks for
$\hat B^{\dagger}_{\mu}$ operators proper for the construction of the
ground state by
\begin{eqnarray}
|\Psi_g\rangle = \prod_{\mu\in I} \hat B^{\dagger}_{\mu} |0\rangle,
\label{e16}
\end{eqnarray}
where $|0\rangle$ is the bare vacuum, $I$ is a set of parameters $\mu$, 
and the $\hat B^{\dagger}_{\mu}$ 
operators satisfy
\begin{eqnarray}
\{ \hat A_{\nu,{\bf i},\sigma} , \hat B^{\dagger}_{\mu} \} =0,
\label{e17}
\end{eqnarray}
for all values of all indices $\nu,\mu,\sigma$ and ${\bf i}$. The motivation
for this step is that given by the requirement (\ref{e17}), for the
wave 
vector $|\Psi \rangle = \prod_{\mu} \hat B^{\dagger}_{\mu} |0\rangle$,
one has the
property
\begin{eqnarray}
\sum_{{\bf i}=1}^{N_c} \sum_{\sigma}\sum_{\nu=1}^{4}
\hat A^{\dagger}_{\nu,{\bf i},\sigma} \hat A_{\nu,{\bf i},\sigma}|\Psi\rangle
=0 .
\label{e18}
\end{eqnarray}
Indeed, (\ref{e18}) holds since starting from (\ref{e17}), the
$\hat A_{\nu,{\bf i},\sigma}$ operators in (\ref{e18}) can be pushed in
front of the vacuum state, hence given by 
$\hat A_{\nu,{\bf i},\sigma}|0\rangle=0$, the equality in Eq.(\ref{e18}) is
satisfied.
 
In the second step ii) one chooses from all the possible $\hat B^{\dagger}_{
\mu}$ operators those $\hat B^{\dagger}_{\mu_j}$,
$I=\{\mu_1,\mu_2, ...,\mu_j, ..., \mu_M\}$, which satisfy also $\hat H_U (
\prod_{\mu\in I} \hat B^{\dagger}_{\mu} |0\rangle) =0$.

\subsection{The operators needed for the construction of the ground state}

\subsubsection{Requirements for the operators building up the ground state}

One considers the $\hat B^{\dagger}_{\mu}$ operators as the most general 
linear combination of creation operators with fixed spin projection acting
on each lattice site of the system. Consequently, besides a fixed $\mu=\mu_1$
index, also the spin projection $\sigma$ will be explicitly used as a label, 
and one has
\begin{eqnarray}
\hat B^{\dagger}_{\mu_1,\sigma}= \sum_{{\bf j}} (x_{1,{\bf j}} 
\hat c^{\dagger}_{{\bf j}+{\bf r}_1,\sigma} +  x_{2,{\bf j}} 
\hat c^{\dagger}_{{\bf j}+{\bf r}_2,\sigma} +  x_{3,{\bf j}} 
\hat c^{\dagger}_{{\bf j}+{\bf r}_3,\sigma} +  x_{4,{\bf j}} 
\hat c^{\dagger}_{{\bf j}+{\bf r}_4,\sigma}),
\label{e19}
\end{eqnarray}
where the site ${\bf j}$ runs over all lattice sites of the sublattice $S_1$,
and the $x_{\nu,i}$ coefficients represent the unknown variables which
must be deduced.
The placement of the coefficients $x_{\nu,i}$ is presented in Fig.7.

\begin{figure}[t!]
\epsfxsize=8cm
\centerline{\epsfbox{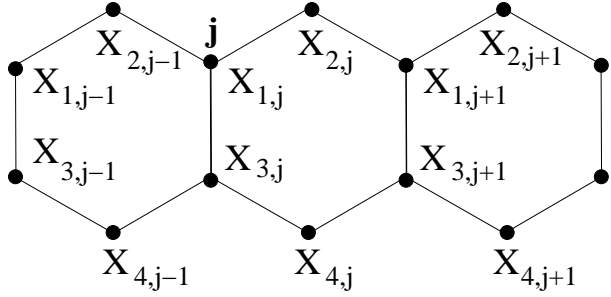}}
\caption{The coefficients $x_{\nu,i}$ from (\ref{e19})
of the sites of the hexagon chain
presented in the neighborhood of the lattice site ${\bf j}$.}
\label{fig7}
\end{figure}

Now using (\ref{e15},\ref{e19}) in (\ref{e17}), one finds the following
linear equations for the $x_{\nu,i}$ coefficients from (\ref{e19}) defined
for the cell constructed at the site ${\bf i}={\bf j}$
\begin{eqnarray}
&&\sqrt{t_e}(x_{2,j}+x_{2,j-1}) = -(\frac{t}{\sqrt{t_e}} x_{1,j} + 
\frac{t_1\sqrt{t_e}}{2t} x_{3,j}),
\nonumber\\
&&\sqrt{t_e}(x_{4,j}+x_{4,j-1}) = -(\frac{t_1\sqrt{t_e}}{2t} x_{1,j} + 
\frac{t}{\sqrt{t_e}}  x_{3,j}),
\nonumber\\
&&\sqrt{t'_1} (x_{1,j}+x_{1,j+1}) = -Q x_{4,j},
\nonumber\\
&&\sqrt{t'_1} (x_{3,j}+x_{3,j+1}) = -Q x_{2,j},
\label{e20}
\end{eqnarray}
where the notation $Q=(2tt'-t_1t_e)/(2t\sqrt{t'_1})$ has been introduced.
From (\ref{e20}) one finds
\begin{eqnarray}
&&x_{1,j+1} = (\frac{Qt_1}{2t\sqrt{t'_1}}-1) x_{1,j}+\frac{Qt}{t_e\sqrt{t'_1}}
x_{3,j}+\frac{Q}{\sqrt{t'_1}}x_{4,j-1},
\nonumber\\
&&x_{3,j+1} = (\frac{Qt_1}{2t\sqrt{t'_1}}-1) x_{3,j}+\frac{Qt}{t_e\sqrt{t'_1}}
x_{1,j}+\frac{Q}{\sqrt{t'_1}}x_{2,j-1},
\nonumber\\
&&x_{2,j}=-\frac{t}{t_e} x_{1,j} -\frac{t_1}{2t}x_{3,j} - x_{2,j-1},
\nonumber\\
&&x_{4,j}=-\frac{t_1}{2t} x_{1,j} -\frac{t}{t_e}x_{3,j} - x_{4,j-1}.
\label{e21}
\end{eqnarray}
The equations from (\ref{e21}) can be written in a matrix form as
\begin{eqnarray}
{\bf Z}_j = \tilde {\bf R} \: \: {\bf Z}_{j-1} ,
\label{e22}
\end{eqnarray}
where ${\bf Z}_j$ is a column vector with 4 components, which taken in
order are given by
$x_{1,j+1},x_{3,j+1}, x_{2,j},x_{4,j}$. Similarly, ${\bf Z}_{j-1}$ is the 
column vector with elements $x_{1,j},x_{3,j},x_{2,j-1},x_{4,j-1}$
\begin{eqnarray}
{\bf Z}_j =
\left( \begin{array}{c}
x_{1,j+1}\\
x_{3,j+1}\\
x_{2,j}\\
x_{4,j}\\ 
\end{array} \right), 
\quad
{\bf Z}_{j-1} =
\left( \begin{array}{c}
x_{1,j}\\
x_{3,j}\\
x_{2,j-1}\\
x_{4,j-1}\\ 
\end{array} \right) .
\label{e23}
\end{eqnarray}

Finally, the $4\times 4$ matrix $\tilde {\bf R}$ is defined by

\begin{eqnarray}
\tilde {\bf R} =
\left( \begin{array}{cccc}
\frac{Qt_1}{2t\sqrt{t'_1}}-1 & \frac{Qt}{t_e\sqrt{t'_1}} & 0
& \frac{Q}{\sqrt{t'_1}} \\
\frac{Qt}{t_e\sqrt{t'_1}} &\frac{Qt_1}{2t\sqrt{t'_1}}-1  &  
\frac{Q}{\sqrt{t'_1}} & 0 \\
-\frac{t}{t_e} & -\frac{t_1}{2t} & -1 & 0 \\
-\frac{t_1}{2t} & - \frac{t}{t_e} & 0 & -1 \\
\end{array} \right) .
\label{e24}
\end{eqnarray}
Starting from (\ref{e24}), in the following subsection one shows how it is
possible to construct one (starting) $\hat B^{\dagger}_{\mu_1,\sigma}$
operator.

\subsubsection{The construction of one starting  $\hat B^{\dagger}_{\mu_1,
\sigma}$ operator}

Based on (\ref{e22}) one finds
\begin{eqnarray}
{\bf Z}_{j+m}= \tilde {\bf W}_m {\bf Z}_j, \quad \tilde {\bf W}_m = 
(\tilde {\bf R})^m .
\label{e25}
\end{eqnarray}
Consequently, since periodic boundary conditions are taken into account
(i.e. after a finite number of $m$ steps, in ${\bf Z}_{j+m}$ one must recovers
${\bf Z}_j$), an operator $\hat B^{\dagger}_{\mu_1,\sigma}$ from (\ref{e19}),
proper for the construction of the ground state, is obtained when
the matrix $\tilde {\bf W}_m$ has at least one eigenvalue $1$. In this case
the $x_{\alpha,i}$ unknown parameters from (\ref{e19}) are given exactly by 
the corresponding ${\bf Z}^e_j$ eigenvector
\begin{eqnarray}
\tilde {\bf W}_m {\bf Z}^e_j = {\bf Z}^e_j .
\label{e26}
\end{eqnarray}
One finds in this manner the starting coefficients $x_{\alpha,i}$ in the form
\begin{eqnarray}
x_{1,j}=x^e_{1,j},\quad x_{3,j}=x^e_{3,j},\quad x_{2,j-1}= x^e_{2,j-1},
\quad x_{4,j-1}=x^e_{4,j-1} .
\label{e27}
\end{eqnarray}
For the next $(m-1)$ cells ${\bf Z}_{j'}$, the coefficients 
$x_{1,j'},x_{3,j'},x_{2,j'-1},x_{4,j'-1}$ are obtained by 
successive application of the $\tilde {\bf R}$ matrix to ${\bf Z}^e_j$ as 
follows: $x_{1,j+1},x_{3,j+1},x_{2,j},x_{4,j}$ are obtained from ${\bf Z}_1$ 
deduced as ${\bf Z}_1=\tilde {\bf R} {\bf Z}^e_j$;   
$x_{1,j+2},x_{3,j+2},x_{2,j+1},x_{4,j+1}$ are obtained from  ${\bf Z}_2$ 
deduced as ${\bf Z}_2=(\tilde {\bf R})^2 {\bf Z}^e_j$; etc.;
$x_{1,j+m-1},x_{3,j+m-1},x_{2,j+m-2},x_{4,j+m-2}$ are obtained from  
${\bf Z}_{m-1}$ deduced as ${\bf Z}_{m-1}=(\tilde {\bf R})^{m-1} {\bf Z}^e_j$.
Given by (\ref{e26}), for the coefficients $x_{1,j+m},x_{3,j+m},x_{2,j+m-1},
x_{4,j+m-1}$ entering in ${\bf Z}_m=(\tilde {\bf R})^{m} {\bf Z}^e_j$,
the values $x_{1,j+m}=x_{1,j}, x_{3,j+m}=x_{3,j},
x_{2,j+m-1}=x_{2,j-1}, x_{4,j+m-1}=x_{4,j-1}$ are reobtained. After this step,
starting from the coefficients  $x_{1,j+m+1},x_{3,j+m+1},x_{2,j+m},
x_{4,j+m}$ present in ${\bf Z}_{m+1}$, the expressions are periodically 
repeated
since ${\bf Z}_{m+1}= \tilde {\bf R} {\bf Z}^e_j$ holds again. In this manner a
$\hat B^{\dagger}_{\mu_1,\sigma}$ operator has been constructed for a system
characterized by $N_c= p \times m$, where $p$ is an arbitrary positive integer.

\subsubsection{Generation of new $\hat B^{\dagger}_{\mu_{\alpha},\sigma}$
operators}

Once a $\hat B^{\dagger}_{\mu_1,\sigma}$ operator has been deduced following 
the procedure described above, one has a proper operator for the construction 
of the ground state that has the obtained $x_{\alpha,i}$ coefficients 
structure holding a periodicity of $m$ ${\bf Z}$ cells, the starting cell
being the ${\bf j}$th cell ${\bf Z}_j$. If one moves the starting cell to the
${\bf Z}_{j+1}$ cell, one obtains a new, (usually) linearly independent 
$\hat B^{\dagger}_{\mu_2,\sigma}$ operator. Repeating the procedure, starting 
from $\hat B^{\dagger}_{\mu_1,\sigma}$, maximum $(m-1)$ new operators
$\hat B^{\dagger}_{\mu_{\beta},\sigma}$, $\beta=2,3,4,...,m$ can be created
(the linear independence must be checked at each step).
If more than one unity eigenvalues are present, the procedure can be repeated 
for each eigenvalue, constructing in this manner for the fixed $m$ periodicity,
maximum $q\times m$ different $\hat B^{\dagger}_{\mu,\sigma}$ operators, 
where $q$ represents the number of
unity eigenvalues corresponding to eigenvectors with non-zero norm.
An exemplification is presented in the Appendix. The procedure can then be 
repeated for another $m$.

\subsubsection{The ground state wave function}

Using the operators deduced above, the $|\Psi\rangle$ wave vector described
below (\ref{e17}) has the form
\begin{eqnarray}
|\Psi \rangle =\prod_{\mu} \prod_{\sigma_{\mu}} 
\hat B^{\dagger}_{\mu,\sigma_{\mu}} |0\rangle .
\label{e28}
\end{eqnarray}
One notes that the $\hat B^{\dagger}_{\mu,\sigma}$ operators (given by 
(\ref{e22}))
are all extended operators. Hence increasing the number of carriers 
and taking into account that all $\hat B^{\dagger}_{\mu,\sigma}$ operators are
confined in the same quasi 1D structure, it results that connectivity 
conditions
will be satisfied between different $\hat B^{\dagger}_{\mu,\sigma}$ operators
with different $\mu$ indices (e.g. $\hat B^{\dagger}_{\mu_i,\sigma}$ and
$\hat B^{\dagger}_{\mu_j,\sigma}$ operators for $i \ne j$ will touch each 
other, hence will act on a given finite number of common sites). 
For the example provided in the Appendix A. this happens starting from
the $N=2$ number of particles. Consequently, in order to minimize the ground 
state energy, all $\sigma$ indices in (\ref{e28}) will be fixed to the same 
value ($\sigma_{\mu}=\sigma_{\mu'}$ for all $\mu,\mu'$) providing in this 
manner zero double occupancy, hence a ferromagnetic ground state of the form
\begin{eqnarray}
|\Psi_g \rangle =\prod_{\mu}  
\hat B^{\dagger}_{\mu,\sigma} |0\rangle ,
\label{e29}
\end{eqnarray} 
where $\sigma$ is fixed. Indeed $|\Psi_g\rangle$ from (\ref{e29}), as 
constructed from (\ref{e28}), will satisfy $(\sum_{{\bf i}=1}^{N_c} 
\sum_{\sigma}\sum_{\nu=1}^{4}\hat A^{\dagger}_{\nu,{\bf i},\sigma} 
\hat A_{\nu,{\bf i},\sigma}) |\Psi_g\rangle =0$ for the first part of
the transformed Hamiltonian from (\ref{e3}), while $\hat H_U |\Psi_g\rangle
=0$ will be given by the absence of the double occupancy. Consequently,
$|\Psi_g\rangle$ represents the ground state of $\hat H$ from (\ref{e3}),
the ground state energy being $E_g= KN$, where $K$ is given in (\ref{e13}).
Since one $\hat B^{\dagger}_{\mu,\sigma}$ operator introduces in fact one
particle into the system, the particle number (hence the concentration of 
carriers) at which $|\Psi_g\rangle$ in (\ref{e29}) is defined, is given by
the number of components of the product from (\ref{e29}). 

As described, $|\Psi_g\rangle$ represents a ferromagnetic state. This ground
state is itinerant and metallic because of the following reasons: i) the
ground state wave function is extended, and  ii) up to 
the $N_{max}$ value equal to the
number of components of the product from (\ref{e29}),  one has for
the particle number ($N < N_{max}$) dependent chemical potential $\mu_c$, 
the expression
$\delta \mu_c = \mu_c(+)-\mu_c(-)=0$ where $\mu_c(+)=E_g(N+1)-E_g(N)$ and
$\mu_c(-)=E_g(N)-E_g(N-1)$, see also Ref.[\cite{lieb}].

Our up to date results show that the ferromagnetism emerges in the low
concentration domain, a pedagogical example being presented in Appendix A.
If one characterizes the carrier concentration by $n=N/N_c$ (i.e. average 
electron 
number per cell), the exemplified case in Appendix A. of six explicitly given 
operators describing
through (\ref{e29}) six interacting electrons, even for a chain made of 36 
cells (for example) 
gives $n=1/6$, a
far from zero finite concentration value. Consequently, the reported results 
clearly
demonstrate the presence of itinerant ferromagnetism in the systems under 
study at low concentration, for
finite chains (treated with periodic boundary conditions). Since 36 cells is 
a huge number for
oligo-acenes (i.e. {\it few} linearly fused hexagons) placed in the center of 
the attention \cite{BB13},
our findings represent genuine information for potential application
possibilities. 

Furthermore, one underlines
that since (as shown in Sec.III.)
flat bands are not present in the system, the here described ferromagnetism 
is not of
flat  band type, and is provided by a joint effect of correlation and 
confinement. 

One notes that for the finite sample case explicitly exemplified in
Appendix A, the Hubbard repulsion must satisfy only 
the $U > 0$ condition,
the $U$ value itself being without importance. The phase diagram region where 
the presented solution
occurs is not severely restricted. Besides two sign requirements present for 
two hopping matrix elements
($t_e,t'_1 > 0$), only one condition for on-site one-particle potentials is 
present (see (\ref{e14})),
which requires a fixed value for $\epsilon_0$ acting on external sites. 
Since this on-site potential
can be tuned by an external gate potential \cite{BB2a}, the described 
itinerant ferromagnetism
can be in principle even switched on by an external electric field.

A detailed study of the phase diagram and the behavior in the thermodynamic 
limit is in preparation, and will be published elsewhere.

\section{Summary and conclusions}

A zig-zag hexagon chain described by a Hubbard type of model containing 
on-site Coulomb repulsion is
analyzed in exact terms by a technique based on positive semidefinite 
operators. The calculations
effectuated with periodic boundary conditions provide exact ground states 
of ferromagnetic and itinerant
nature in the low concentration region. Flat bands are not present in the
non-interacting band structure, and the ferromagnetism is created by a 
joint effect of correlations and confinement. The parameter space region 
where the described phase emerges is not severely restricted,
and the unique on-site one-particle potential which must have a fixed value 
can in principle be tuned by external gate potentials.  

The reason for the emergence of ferromagnetism is connected to two aspects. 
First, all operators
entering in the construction of the ground state turn out to be extended. 
This is an interesting property of the studied system
since it has not been observed for triangle, quadrilateral, and
pentagon chains \cite{BB7,BB8,BB12c}. Second, all these extended operators 
with contributions in each cell 
act on lattice sites confined in the chain structure under consideration.
Hence by increasing the carrier concentration, connectivity conditions 
between the operators of the
ground state wave vector will emerge. These connectivity conditions 
(i.e. different operators
describing different electrons act on common lattice sites) fix the spin 
indices to the same value,
minimizing in this manner as much as possible the double occupancy in order 
to reduce the ground state
energy, and provide the ferromagnetic ground state. The itinerant
nature of the obtained ferromagnetic ground state may open 
new application possibilities in spintronics.

One adds below some observations related to the exact nature of the results.
One knows that the flat band, or nearly flat band ferromagnetism can be
described in exact terms \cite{BB10a,BB10b}. Furthermore, it was shown
that ferromagnetism
emerging for completely dispersive band structure provided by spin-spin 
interactions in itinerant systems can also be described in exact terms
\cite{BBu}. The here reported results enlarge the spectrum of exact
descriptions relating ferromagnetism in completely dispersive systems: one
demonstrates that when the joint action of confinement and correlations is
present, itinerant ferromagnetism can be described in exact terms at finite
value of the interaction even in the frame of a non-extended Hubbard type 
of model.  

The deduced results deserve a further observation as well:
In the search for organic ferromagnetism is no more necessary 
to look for hopping and on-site one-particle potential values in 
different chain 
structures in order to introduce flat bands in the system leading to flat 
band ferromagnetism. Itinerant ferromagnetism can be obtained 
even in completely dispersive  systems, for example 
by a joint effect of correlations
and confinement.

\section*{Acknowledgements}

Z.G. gratefully acknowledges
financial support provided by the Hungarian Research Fund through
Contracts No. OTKA-T48782 and the Alexander von Humboldt Foundation.

\appendix

\section{Exemplification for the construction of 
$\hat B^{\dagger}_{\mu,\sigma}$ operators}

Let us consider the following pedagogical example
for the construction of operators needed in the ground state wave
vector. Using the notations
$a=Q/\sqrt{t'_1}, b=t/t_e, c=t_1/(2t)$, 
the $\tilde {\bf R}$ matrix from
(\ref{e24}) becomes
\begin{eqnarray}
\tilde {\bf R} =
\left( \begin{array}{cccc}
ac-1 & ab & 0 & a \\
ab & ac-1  & a  & 0 \\
-b & -c & -1 & 0 \\
-c & -b & 0 & -1 \\
\end{array} \right) .
\label{ea1}
\end{eqnarray}
Using for simplicity the conditions $c=-b, a=-1/b$ \cite{obs1}, 
for $\tilde {\bf W}_4$
one obtains four unity eigenvalues. One of eigenvectors has only zero
elements (i.e. zero norm), so this cannot be used to construct physical 
states, but
three other eigenvectors have non-zero norm. These are the following ones
\begin{eqnarray}
{\bf Z}^e_1 =
\left( \begin{array}{c}
0\\
0\\
-1\\
1\\ 
\end{array} \right), 
\quad
{\bf Z}^e_2 =
\left( \begin{array}{c}
0\\
1\\
0\\
0\\ 
\end{array} \right) ,
\quad
{\bf Z}^e_3 =
\left( \begin{array}{c}
1\\
0\\
0\\
0\\ 
\end{array} \right) .
\label{ea2}
\end{eqnarray}

Let us consider first the ${\bf Z}^e_1$ eigenvector. For it one has
\begin{eqnarray}
\tilde {\bf R} {\bf Z}^e_1 =
\left( \begin{array}{c}
-\frac{1}{b}\\
\frac{1}{b}\\
1\\
-1\\ 
\end{array} \right), 
\quad
{\tilde {\bf R}}^2 {\bf Z}^e_1 =
\left( \begin{array}{c}
0\\
0\\
1\\
-1\\ 
\end{array} \right), 
\quad
{\tilde {\bf R}}^3 {\bf Z}^e_1 =
\left( \begin{array}{c}
\frac{1}{b}\\
-\frac{1}{b}\\
-1\\
1\\ 
\end{array} \right), 
\quad
{\tilde {\bf R}}^4 {\bf Z}^e_1 =
\left( \begin{array}{c}
0\\
0\\
-1\\
1\\ 
\end{array} \right). 
\label{ea3}
\end{eqnarray}
Taking ${\bf Z}_j={\bf Z}^e_1$ one finds based on (\ref{ea3}) the following 
starting $\hat B^{\dagger}_{\mu_1,\sigma}$ operator
\begin{eqnarray}
\hat B^{\dagger}_{\mu_1,\sigma} &=& ... + (- c^{\dagger}_{{\bf j}+{\bf r}_2,
\sigma}
+ c^{\dagger}_{{\bf j}+{\bf r}_4,\sigma}) + (- \frac{1}{b} c^{\dagger}_{
{\bf j}+2{\bf a}+{\bf r}_1,\sigma} +\frac{1}{b} c^{\dagger}_{{\bf j}+2{\bf a}
+{\bf r}_3,\sigma} + c^{\dagger}_{{\bf j}+{\bf a}+{\bf r}_2,\sigma} -
c^{\dagger}_{{\bf j}+{\bf a}+{\bf r}_4,\sigma}) 
\nonumber\\
&+& (c^{\dagger}_{{\bf j}+
2{\bf a}+{\bf r}_2,\sigma} - c^{\dagger}_{{\bf j}+2{\bf a}+{\bf r}_4,\sigma}) 
+ (\frac{1}{b} c^{\dagger}_{{\bf j}+4{\bf a}+{\bf r}_1,\sigma} - 
\frac{1}{b} c^{\dagger}_{{\bf j}+4{\bf a} + {\bf r}_3,\sigma} - 
c^{\dagger}_{{\bf j}+3{\bf a}+{\bf r}_2,\sigma} + c^{\dagger}_{{\bf j}+
3{\bf a}+{\bf r}_4,\sigma}) 
\nonumber\\
&+& (- c^{\dagger}_{{\bf j}+4{\bf a}+{\bf r}_2,
\sigma} +  c^{\dagger}_{{\bf j}+4{\bf a}+{\bf r}_4,\sigma}) + ... ,
\label{ea4}
\end{eqnarray}
where  the contributions from fixed ${\bf Z}$ cells are collected in 
parentheses.
The second $\hat B^{\dagger}_{\mu_2,\sigma}$ operator is obtained from
(\ref{ea4}) by translating the starting point of the period with one cell, 
i.e. taking ${\bf Z}_{j+1}={\bf Z}^e_1$. One gets
\begin{eqnarray}
\hat B^{\dagger}_{\mu_2,\sigma} &=& ... + (\frac{1}{b} c^{\dagger}_{{\bf j}
+ {\bf a}+{\bf r}_1,\sigma} - \frac{1}{b} c^{\dagger}_{{\bf j}+ {\bf a} 
+ {\bf r}_3,\sigma} - c^{\dagger}_{{\bf j}+{\bf r}_2,\sigma} + c^{\dagger}_{
{\bf j}+ {\bf r}_4,\sigma}) +  (- c^{\dagger}_{{\bf j}+{\bf a}+{\bf r}_2,
\sigma} + c^{\dagger}_{{\bf j}+{\bf a}+{\bf r}_4,\sigma}) 
\nonumber\\
&+& (- \frac{1}{b} c^{\dagger}_{
{\bf j}+3{\bf a}+{\bf r}_1,\sigma} +\frac{1}{b} c^{\dagger}_{{\bf j}+3{\bf a}
+{\bf r}_3,\sigma} + c^{\dagger}_{{\bf j}+2{\bf a}+{\bf r}_2,\sigma} -
c^{\dagger}_{{\bf j}+2{\bf a}+{\bf r}_4,\sigma}) +  (c^{\dagger}_{{\bf j}+
3{\bf a}+{\bf r}_2,\sigma} - c^{\dagger}_{{\bf j}+3{\bf a}+{\bf r}_4,\sigma})
\nonumber\\
&+&(\frac{1}{b} c^{\dagger}_{{\bf j}+5{\bf a}+{\bf r}_1,\sigma} - 
\frac{1}{b} c^{\dagger}_{{\bf j}+5{\bf a} + {\bf r}_3,\sigma} - 
c^{\dagger}_{{\bf j}+4{\bf a}+{\bf r}_2,\sigma} + c^{\dagger}_{{\bf j}+
4{\bf a}+{\bf r}_4,\sigma}) + ... . 
\label{ea5}
\end{eqnarray}
A further translation of ${\bf Z}^e_1$ to the following cell by taking
${\bf Z}_{j+2}={\bf Z}^e_1$ provides $-\hat B^{\dagger}_{\mu_1,\sigma}$ while
the fourth translation by taking ${\bf Z}_{j+3}={\bf Z}^e_1$ reproduces
$\hat B^{\dagger}_{\mu_1,\sigma}$. Hence these steps do not provide new 
linearly independent $\hat B^{\dagger}_{\mu,\sigma}$ operators.

Now one concentrates on the ${\bf Z}^e_2$ eigenvector. In this case one
obtains
\begin{eqnarray}
\tilde {\bf R} {\bf Z}^e_2 =
\left( \begin{array}{c}
-1\\
0\\
b\\
-b\\ 
\end{array} \right), 
\quad
{\tilde {\bf R}}^2 {\bf Z}^e_2 =
\left( \begin{array}{c}
1\\
0\\
0\\
0\\ 
\end{array} \right), 
\quad
{\tilde {\bf R}}^3 {\bf Z}^e_2 =
\left( \begin{array}{c}
0\\
-1\\
-b\\
b\\ 
\end{array} \right), 
\quad
{\tilde {\bf R}}^4 {\bf Z}^e_2 =
\left( \begin{array}{c}
0\\
1\\
0\\
0\\ 
\end{array} \right). 
\label{ea6}
\end{eqnarray}
Taking ${\bf Z}_j={\bf Z}^e_2$ one obtains the third linearly independent 
$\hat B^{\dagger}_{\mu,\sigma}$
operator as follows
\begin{eqnarray}
\hat B^{\dagger}_{\mu_3,\sigma} &=& ... + (\hat c^{\dagger}_{{\bf j}+{\bf a}
+{\bf r}_3,\sigma} ) + (-\hat c^{\dagger}_{{\bf j}+2{\bf a}+{\bf r}_1,\sigma} 
+ b \hat c^{\dagger}_{{\bf j}+{\bf a}+{\bf r}_2,
\sigma}-b \hat c^{\dagger}_{{\bf j}+{\bf a}+{\bf r}_4,\sigma}) 
\nonumber\\
&+& (\hat c^{\dagger}_{{\bf j}+ 3{\bf a}+{\bf r}_1,\sigma} ) 
+ (- \hat c^{\dagger}_{{\bf j}+4{\bf a}+{\bf r}_3,\sigma}  
-b \hat c^{\dagger}_{{\bf j}+3{\bf a} + {\bf r}_2,\sigma}  
+ b \hat c^{\dagger}_{{\bf j}+3{\bf a}+{\bf r}_4,\sigma}) 
\nonumber\\
&+& (\hat c^{\dagger}_{{\bf j}+5{\bf a}+{\bf r}_3,\sigma}) + ... .
\label{ea7}
\end{eqnarray}
The fourth $\hat B^{\dagger}_{\mu_4,\sigma}$ operator is obtained from
(\ref{ea7}) by translating the starting point of the period with one cell, 
i.e. taking ${\bf Z}_{j+1}={\bf Z}^e_2$. One gets
\begin{eqnarray}
\hat B^{\dagger}_{\mu_4,\sigma} &=& ... + (- \hat c^{\dagger}_{{\bf j}
+ {\bf a}+{\bf r}_3,\sigma} - b \hat c^{\dagger}_{{\bf j}+ {\bf r}_2,\sigma} 
+b \hat c^{\dagger}_{{\bf j}+{\bf r}_4,\sigma}) +  (\hat c^{\dagger}_{{\bf j}+
2 {\bf a}+{\bf r}_3,\sigma})
\nonumber\\
&+&
(- \hat c^{\dagger}_{{\bf j}+3{\bf a}+{\bf r}_1,\sigma} + b \hat c^{\dagger}_{
{\bf j}+2{\bf a}+{\bf r}_2,\sigma} - b \hat c^{\dagger}_{{\bf j}+2 {\bf a}+
{\bf r}_4,\sigma}) + (\hat c^{\dagger}_{{\bf j}+4 {\bf a}+{\bf r}_1,\sigma})
\nonumber\\
&+& (- \hat c^{\dagger}_{{\bf j}+5{\bf a}+{\bf r}_3,\sigma} -b
\hat c^{\dagger}_{{\bf j}+4{\bf a}+{\bf r}_2,\sigma} + b \hat c^{\dagger}_{
{\bf j}+4{\bf a}+{\bf r}_4,\sigma}) +  (\hat c^{\dagger}_{{\bf j}+ 6{\bf a}+
{\bf r}_3,\sigma} ) + ... .
\label{ea8}
\end{eqnarray}
Continuing the translation by one cell and taking ${\bf Z}_{j+2}={\bf Z}^e_2$
one obtains the $\hat B^{\dagger}_{\mu_5,\sigma}$ operator as follows
\begin{eqnarray}
\hat B^{\dagger}_{\mu_5,\sigma} &=& ... + (\hat c^{\dagger}_{{\bf j}
+ {\bf a}+{\bf r}_1,\sigma}) + (-\hat c^{\dagger}_{{\bf j}+2{\bf a}+{\bf r}_3,
\sigma} - b \hat c^{\dagger}_{{\bf j}+ {\bf a}+{\bf r}_2,\sigma} 
+b \hat c^{\dagger}_{{\bf j}+{\bf a}+{\bf r}_4,\sigma}) 
\nonumber\\
&+& (\hat c^{\dagger}_{{\bf j}+3{\bf a}+{\bf r}_3,\sigma}) +
(- \hat c^{\dagger}_{{\bf j}+4 {\bf a}+{\bf r}_1,\sigma} + b \hat c^{\dagger}_{
{\bf j}+3{\bf a}+{\bf r}_2,\sigma}  - b \hat c^{\dagger}_{
{\bf j}+3{\bf a}+{\bf r}_4,\sigma}) 
\nonumber\\
&+& (\hat c^{\dagger}_{{\bf j}+5{\bf a}+{\bf r}_1,\sigma}) +
(- \hat c^{\dagger}_{{\bf j}+6{\bf a}+{\bf r}_3,\sigma} - b \hat c^{\dagger}_{
{\bf j}+5{\bf a}+{\bf r}_2,\sigma} + b \hat c^{\dagger}_{{\bf j}+5 {\bf a}+
{\bf r}_4,\sigma}) + ... .
\label{ea9}
\end{eqnarray}
With ${\bf Z}_{j+3}={\bf Z}^e_2$ one obtains the last independent
operator generated by  ${\bf Z}^e_2$, namely
$\hat B^{\dagger}_{\mu_6,\sigma}$, in the form
\begin{eqnarray}
\hat B^{\dagger}_{\mu_6,\sigma} &=& ... + (- \hat c^{\dagger}_{{\bf j}
+ {\bf a}+{\bf r}_1,\sigma} + b \hat c^{\dagger}_{{\bf j}+ {\bf r}_2,\sigma} 
-b \hat c^{\dagger}_{{\bf j}+{\bf r}_4,\sigma}) +  (\hat c^{\dagger}_{{\bf j}+
2 {\bf a}+{\bf r}_1,\sigma})
\nonumber\\
&+&
(- \hat c^{\dagger}_{{\bf j}+3{\bf a}+{\bf r}_3,\sigma} - b \hat c^{\dagger}_{
{\bf j}+2{\bf a}+{\bf r}_2,\sigma} + b \hat c^{\dagger}_{{\bf j}+2 {\bf a}+
{\bf r}_4,\sigma}) + (\hat c^{\dagger}_{{\bf j}+4 {\bf a}+{\bf r}_3,\sigma})
\nonumber\\
&+& (- \hat c^{\dagger}_{{\bf j}+5{\bf a}+{\bf r}_1,\sigma} + b
\hat c^{\dagger}_{{\bf j}+4{\bf a}+{\bf r}_2,\sigma} - b \hat c^{\dagger}_{
{\bf j}+4{\bf a}+{\bf r}_4,\sigma}) +  (\hat c^{\dagger}_{{\bf j}+ 6{\bf a}+
{\bf r}_1,\sigma} ) + ... .
\label{ea10}
\end{eqnarray}
The use of ${\bf Z}^e_3$ leads to $\hat B^{\dagger}_{\mu,\sigma}$ operators 
linearly dependent on $\hat B^{\dagger}_{\mu_3,\sigma},\hat B^{\dagger}_{
\mu_4,\sigma},..., \hat B^{\dagger}_{\mu_6,\sigma}$.

Further operators $\hat B^{\dagger}_{\mu}$ from the eigenvectors of 
$\tilde {\bf W}_4$ presented in (\ref{ea2}) cannot be constructed. But
the list of new independent $\hat B^{\dagger}_{\mu}$ operators can be 
continued by looking for unity eigenvalues of $\tilde {\bf W}_n$, $n \ne 4$,
and continuing the procedure presented above. For example in the present case
$\tilde {\bf W}_2$ has also unity eigenvalue holding eigenvector with non-zero
norm ${\bf Z}^e_4$ with properties
\begin{eqnarray}
{\bf Z}^e_4 =
\left( \begin{array}{c}
1\\
1\\
0\\
0\\ 
\end{array} \right), 
\quad
\tilde {\bf R} {\bf Z}^e_4 =
\left( \begin{array}{c}
-1\\
-1\\
0\\
0\\ 
\end{array} \right) ,
\label{ea11}
\end{eqnarray} 
leading to following contribution by ${\bf Z}_{j}={\bf Z}^e_4$.

The linear independence of the explicit $\hat B^{\dagger}_{\mu,\sigma}$ 
operators present in 
(\ref{ea4},\ref{ea5},\ref{ea7} - \ref{ea10}) results from the fact that
each of them has a specific (and different) group of missing initial 
fermionic creation operators.


\begin{references}
\bibitem{obs1}
These conditions represent only a restriction of the parameter space into
a domain where the example of the Appendix A is present.
\end{references}
\end{document}